\documentclass{article}

\usepackage{microtype}
\usepackage{graphicx}
\usepackage{subfigure}
\usepackage{booktabs} 

\usepackage{hyperref}

\usepackage{makecell}
\usepackage{tabularx}

\usepackage[accepted]{icml2023}

\usepackage{amsmath}
\usepackage{amssymb}
\usepackage{mathtools}
\usepackage{amsthm}
\usepackage{bm} 

\usepackage[capitalize,noabbrev]{cleveref}

\theoremstyle{plain}

\theoremstyle{definition}

\theoremstyle{remark}

\usepackage[textsize=tiny]{todonotes}

\icmltitlerunning{Siamese SIREN: Audio Compression with Implicit Neural Representations}

\begin{document}

\twocolumn[
\icmltitle{Siamese SIREN: Audio Compression with Implicit Neural Representations}



\icmlsetsymbol{equal}{*}

\begin{icmlauthorlist}
\icmlauthor{Luca A. Lanzend\"orfer}{sch}
\icmlauthor{Roger Wattenhofer}{sch}

\end{icmlauthorlist}

\icmlaffiliation{sch}{ETH Zurich, Switzerland}

\icmlcorrespondingauthor{Luca A. Lanzendörfer}{lanzendoerfer@ethz.ch}
\icmlcorrespondingauthor{Roger Wattenhofer}{wattenhofer@ethz.ch}

\icmlkeywords{ICML, Machine Learning, Audio Compression, Implicit Neural Representation, Representation Learning, Neural Compression}

\vskip 0.3in
]



\printAffiliationsAndNotice{}  

\begin{abstract}
Implicit Neural Representations (INRs) have emerged as a promising method for representing diverse data modalities, including 3D shapes, images, and audio. While recent research has demonstrated successful applications of INRs in image and 3D shape compression, their potential for audio compression remains largely unexplored. Motivated by this, we present a preliminary investigation into the use of INRs for audio compression. Our study introduces Siamese SIREN, a novel approach based on the popular SIREN architecture. Our experimental results indicate that Siamese SIREN achieves superior audio reconstruction fidelity while utilizing fewer network parameters compared to previous INR architectures.
\end{abstract}

\section{Introduction}
\label{introduction}

INRs have become known as an alternative representation for 3D shapes \cite{Park2019DeepSDFLC, mescheder2019occupancy}, and have since been successfully applied to other data modalities such as radiance fields, images, and audio \cite{sitzmann2020implicit, Yu2020pixelNeRFNR, chen2021learning, mildenhall2021nerf, zuiderveld2021lightweight, szatkowski2022hypersound}.

In this paper, we apply INRs to audio compression, that is we approximate the audio signal function $f: \mathcal{T} \rightarrow \mathbb{R}$, where $\mathcal{T}$ is the time input domain and $\mathbb{R}$ is the amplitude output domain, with a small neural network. We take inspiration from the recent work on compression with INRs \cite{dupont2021coin, dupont2022coin++, strümpler2022implicit} and build on previous work in audio INR \cite{sitzmann2020implicit, zuiderveld2021lightweight, szatkowski2022hypersound}.

Even though INRs cannot yet compete with other data compression approaches in the visual and audio domain~\cite{dupont2022coin++, strümpler2022implicit}, we believe it still warrants further research. INRs have some interesting properties, such as being resolution-invariant to the input data, meaning the storage size does not scale with the input size, as well as having the ability to reconstruct data using any arbitrary resolution during inference.

However, a significant challenge arises when reconstructing audio with INRs. In images, noise may be present, but is often less noticeable. In audio data, however, even relatively small reconstruction errors become clearly perceivable in the form of stationary background noise due to the logarithmic nature of human hearing (Weber–Fechner law, cf. \cref{appendix:noise-perception}). This noise thus becomes more pronounced the further we reduce model size and quantize model weights.

\begin{figure}[t]
\centering
\includegraphics[width=0.82\linewidth]{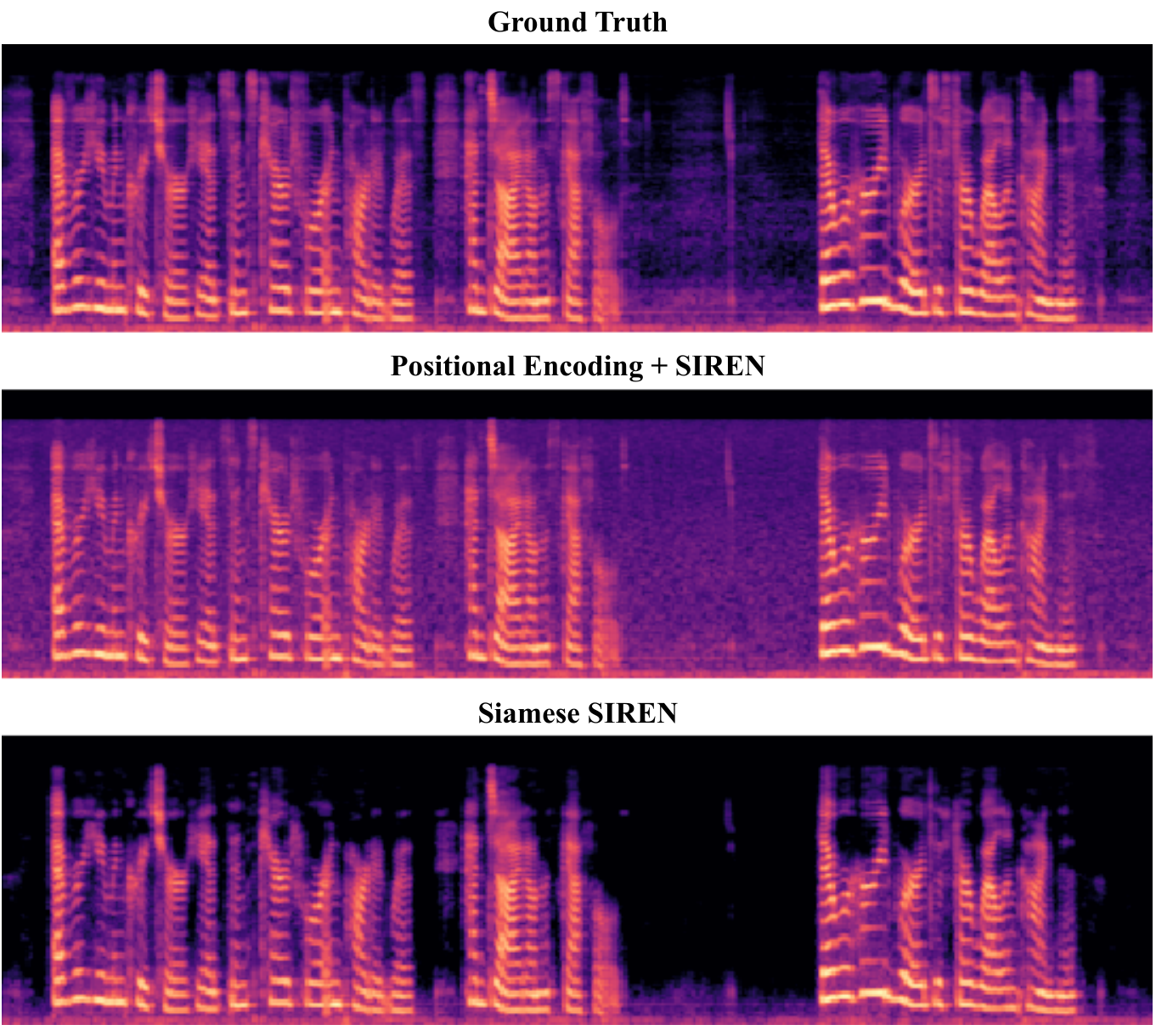}
\caption{Log-mel spectrogram of a random 10-second LibriSpeech sample. We observe background noise produced from SIREN with positional encoding. Siamese SIREN is able to remove the noise by computing the noise estimate.}
\label{fig:siren-siam-comparison}
\end{figure}

The above trade-off can be phrased in the general case as the following Pareto optimization problem: Let $\bm{p}$ be the parameters of a candidate INR, let $\mathcal{D}_f = \left\{ (t, f(t)) : t \in \mathcal{T} \right\}$ be the data of the audio sample $f$, and let $q$ be a quality measure (cf. \cref{experiments}).
Denote the memory footprint by $\left|\,\cdot\,\right|$.
Solve
\begin{align}
    &\max\left( q(\bm{p}, \mathcal{D}_f), \frac{\left|\mathcal{D}_f\right|}{\left|\bm{p}\right|} \right) \\
    &\,\text{subject to } \text{MSE}\left( \bm{p}, \mathcal{D}_f \right) \to 0, \nonumber
\end{align}
where the fraction to be optimized represents the compression ratio achieved by the INR, and where constraint convergence is to come from gradient-descent training of $\bm{p}$ on $\mathcal{D}_f$ until complete convergence (training for overfitting).

To address this problem, we propose Siamese SIREN, an INR model built on top of the general-purpose SIREN architecture \cite{sitzmann2020implicit}. The basic idea of our approach is to add two twin extensions to the standard SIREN model, both extensions trying to approximate the original audio signal $f$. Since both extensions will contain noise, but different noise, their difference can be leveraged to remove the noise from the reconstruction $\hat{f}$. 

We demonstrate the viability of our approach via a set of audio metrics, log-mel spectrograms, and audio samples. The code and examples are available at \href{https://github.com/lucala/siamese-siren/}{https://github.com/lucala/siamese-siren}.

\section{Background}
\label{background}

\textit{INRs} are a class of functions, where one set of function parameters $\bm{p}$ describes one data sample $\mathcal{D}$. In particular, an INR is a neural network trained on $\mathcal{D}$ that approximates $f$.

\textit{Sinusoidal Representation Networks}, referred to as SIREN, are a particular class of INR models \cite{sitzmann2020implicit} that use the multi-layer perceptron (MLP) architecture with sine functions as their activation functions:
\begin{align}
    \phi_i: x_i \rightarrow sin\big(\omega_i\cdot(W_ix_i+b_i)\big),
\end{align}
where $\phi_i$ is the $i^{th}$ layer of the network, $W_i$ and $b_i$ are the weight matrix and bias vector of the $i^{th}$ layer, respectively. The authors found the frequency scaling hyperparameter $\omega_i$ helps SIREN converge faster. They set $\omega_i=30$, with $\omega_0=3000$ in the case of audio.

SIREN INRs have been shown to outperform standard ReLU-activated INRs on images, audio, and 3D geometry \cite{sitzmann2020implicit}.

\textit{Positional Encoding} (PE) has been shown to help INRs learn high-frequency representations \cite{tancik2020fourier, mildenhall2021nerf, Benbarka2021SeeingIN, strümpler2022implicit}. We observe the same effect and also utilize PE, transforming the input into a high-dimensional embedding:
\begin{align}
    \gamma(t) = \big(t, sin(\sigma^0\pi t), cos(\sigma^0\pi t),\dotsc, \nonumber \\
    sin(\sigma^L\pi t), cos(\sigma^L\pi t) \big), 
\end{align}
where $t$ is a normalized point in time, $\sigma$ is a frequency scaling term, and $L$ is the number of frequencies. 

We propose a novel extension of SIREN, which we call \textit{Siamese SIREN}. This architecture is motivated by our finding that small reconstruction errors of the waveform produced by SIREN contain audible background noise, even when training for tens of thousand of iterations (cf. \cref{appendix:siren-audio}). This error is often magnified after quantizing the network weights. To remove the background noise of the reconstructed signal, we use \textit{Noise Reduce} \cite{tim_sainburg_2019_3243139}, an algorithm which computes a spectrogram of the signal and noise estimate. The signal and noise estimate are used to compute a noise threshold for each frequency band. A noise mask is computed based on the threshold, which is in turn used to remove the noise.

\begin{figure}[t]
\includegraphics[width=\linewidth]{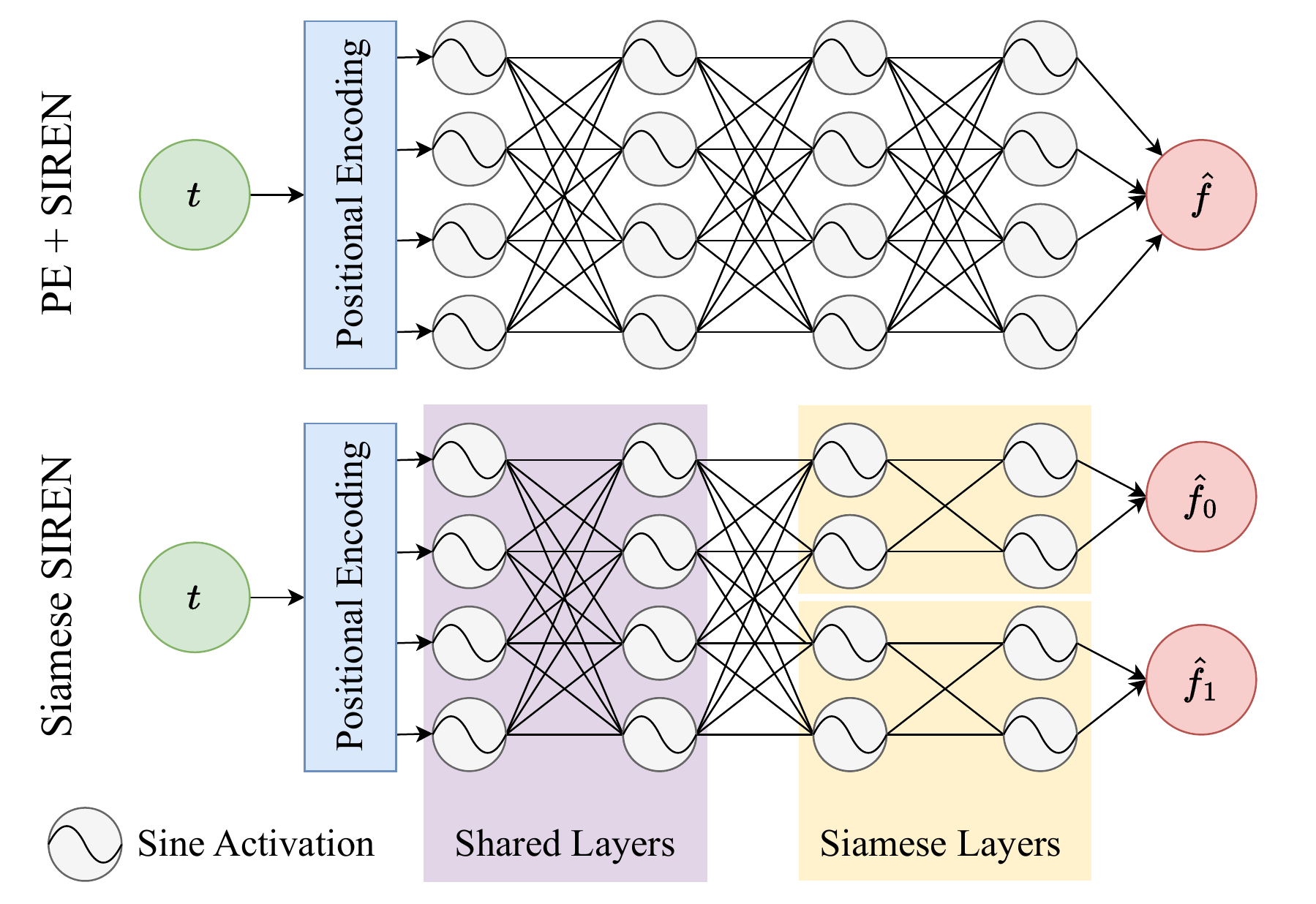}
\caption{Overview of our proposed Siamese SIREN architecture, compared to SIREN with positional encoding (PE+SIREN). The above diagram illustrates a Siamese SIREN with two shared layers and two siamese layers, where each shared layer contains four units and each siamese layer contains two units. Our experiments are centered around two shared layers with 256 units each, and one siamese layer where each siamese head contains 128 units.}
\label{fig:siam-overview}
\end{figure}

\begin{figure*}[t]
\centering
\includegraphics[width=\textwidth]{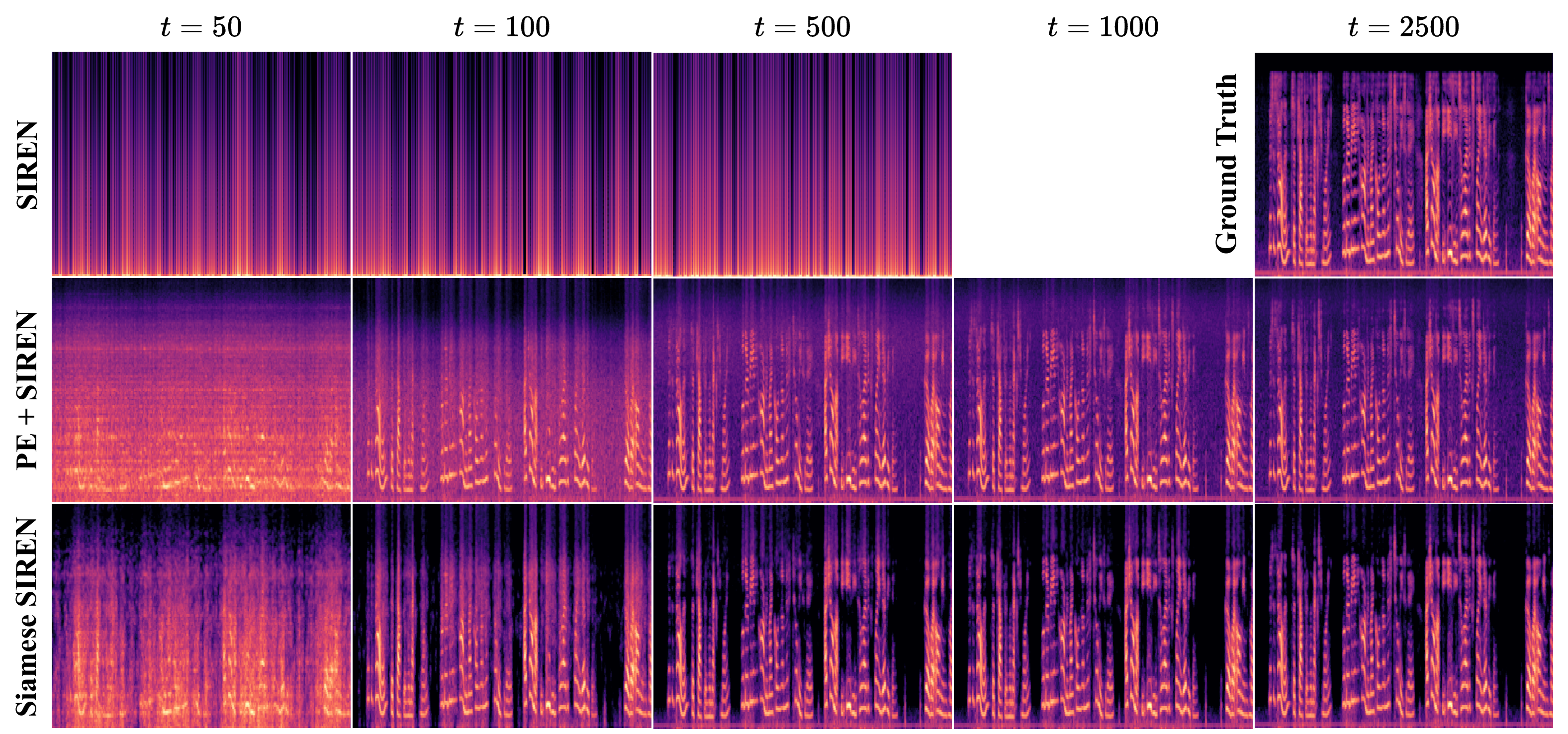}
\caption{Comparison between different quantized models at training iteration $t$ on a 10-second LibriSpeech sample. SIREN without positional encoding (PE) cannot reproduce data after quantization. PE+SIREN is able to reproduce the signal with noise. Siamese SIREN can successfully estimate the background noise and remove it while using less parameters than PE+SIREN.}
\label{fig:training-comparison}
\end{figure*}

To construct a noise estimate for \textit{Noise Reduce}, assume that a noisy reconstructed signal $\hat{f}$ can be linearly decomposed into the true signal $f$ and the noise component $\varepsilon$. Since the distribution from which $\varepsilon$ was sampled is not known in general, it has to be estimated. Training two INRs with different random weight initializiations on the same signal $f$ we obtain two approximations $\hat{f}_0, \hat{f}_1$ of $f$. We use the following rule to arrive at an estimate noise signal $\varepsilon_\bullet$ 
\begin{align}
    \varepsilon_\bullet = \alpha\left(\hat{f}_\bullet - \frac{\hat{f}_0+\hat{f}_1}{2}\right),
\end{align}
where $\alpha$ is a hyperparameter controlling the amplitude of the noise estimate.
We find that tuning $\alpha$ has an impact on results, and we settle at $\alpha=2$ for all experiments.
$\varepsilon_\bullet$ can either be $\varepsilon_0$ or $\varepsilon_1$, and having no preference, we then feed $\varepsilon_0$ as the noise estimate to \textit{Noise Reduce}.

Instead of naively training two INRs to be able to estimate the noise of the signal and thus doubling our parameter count, our proposed Siamese SIREN network merges a subset of layers, reducing the number of required parameters while still allowing for signals $\hat{f}_0$ and $\hat{f}_1$ to be learned. That way, each siamese twin possesses layers that are \textit{shared} as well as layers that are specific only to it (\textit{siamese layers}). In other words, the shared layers form a common backbone for the INR networks, whereas the siamese layers act as two separate heads, see \cref{fig:siam-overview}.

During training, both siamese heads learn to reconstruct the same signal $f$, but due to different random weight initialization of the heads, the reconstructed signals will vary slightly. We leverage this phenomenon to capture the noise estimate that is needed for noise removal, and find it to be effective.

Further on the front of parameter memory footprint reduction, \textit{weight quantization} strategies reduce $\left|\bm{p}\right|$ and increases inference throughput by converting network weights.
These are usually often stored with 32-bit floating point precision, but can be often quantized into smaller data types such as 8-bit integers.
There exist various quantization schemes -- two common approaches are: Post-Training Quantization (PTQ) and Quantization-Aware Training (QAT). QAT tends to achieve lower reconstruction error after quantization. However, QAT needs to either be part of the network while training or fine-tuned after training the unquantized model. PTQ can be applied after training and does not require retraining the network, at the expense of slightly worse reconstruction quality. We use PTQ, as in our early experimentation we found PTQ errors did not significantly affect subjective signal quality.

\section{Experiments}
\label{experiments}

\begin{table*}[t!]
    \centering
    \caption{Comparison between different SIREN configurations after quantization, over random LibriSpeech samples. PE refers to Positional Encoding, $\omega_0$ refers to the scaling factor of the first SIREN layer, $\omega$ refers to all other SIREN layer scaling factors. \textit{1x128} Siamese Layer refers to one layer with 128 units for each siamese head. The original SIREN performs worse since it is not able to reconstruct the signal after weight quantization.}
    \resizebox{\textwidth}{!}{%
    \begin{tabular}{ccccccccccccc}
    \textbf{\makecell{Model\\Name}} & \textbf{\makecell{Shared\\Layers}} & \textbf{\makecell{Siamese\\Layers}} & \textbf{\makecell{PE}} & $\mathbf{\bm{\omega}_0}$ & $\mathbf{\bm{\omega}}$ & \textbf{\#Params ($10^3$)} & \textbf{\makecell{Unquantized\\File Size (kB)}} & \textbf{\makecell{Quantized\\File Size (kB)}} & \textbf{ViSQOL}$\uparrow$ & \textbf{CDPAM}$\downarrow$ & \textbf{PESQ}$\uparrow$ & \textbf{STOI}$\uparrow$ \\
    \hline
    original SIREN & 3x256 & 0 & 0 & 3000 & 30 & 794 & 1594.6 & 410.9 & 1.01 & 0.84 & 1.05 & -0.01 \\ 
    PE + SIREN & 3x256 & 0 & 16 & 30 & 30 & 843 & 1692.9 & 435.5 & 1.63 & 0.2 & 1.91 & 0.93 \\ 
    optimized SIREN & 3x256 & 0 & 16 & 100 & 100 & 843 & 1692.9 & 435.5 & 1.97 & 0.18 & 2.18 & \textbf{0.95} \\
    Siamese SIREN & 2x256 & 1x128 & 16 & 100 & 100 & 513 & 1164.9 & 303.7 & \textbf{2.12} & \textbf{0.16} & \textbf{2.58} & 0.93 \\
    \end{tabular}
    }
    \label{tab:inr-comparison-librispeech}
    \vspace{-1.5em}
\end{table*}

To evaluate the quality of our models, we use the GTZAN \cite{tzanetakis_essl_cook_2001} and LibriSpeech datasets \cite{panayotov2015librispeech}. GTZAN contains 1000 music snippets of ten different genres at 30 seconds each. For speech we use the \texttt{train.100} split of LibriSpeech which contains 14 second audio snippets of English speakers reading passages of text. We crop each audio snippet on the first 10 seconds at a sampling rate of 22050 Hz.

To evaluate the reconstruction quality we mainly rely on ViSQOL \cite{chinen2020visqol}, a metric to determine perceived audio quality. ViSQOL is designed to approximate a subjective listening test and produces Mean Opinion Scores between a reference and a test signal. We also employ CDPAM \cite{manocha2021cdpam}, which approximates perceptual audio similarity between two signals. Additionally, we evaluate our models for LibriSpeech with PESQ \cite{Rix2001PerceptualEO} and STOI \cite{Taal2011AnAF}, which are designed to measure the perceived quality and intelligibility of speech in a signal. See \cref{appendix:training} for more details.

\section{Results}
\label{results}

\begin{table}[h]
    \centering
    \caption{Comparison between different layer configurations over random LibriSpeech samples.}
    \resizebox{\linewidth}{!}{%
    \begin{tabular}{ccccccc}
    \textbf{Shared} & \textbf{Siamese} & \textbf{\#Params ($10^3$)} & \textbf{ViSQOL}$\uparrow$ & \textbf{CDPAM}$\downarrow$ & \textbf{PESQ}$\uparrow$ & \textbf{STOI}$\uparrow$ \\
    \hline
    3x256 & 0 & 843 & 1.5 & 0.23 & 2.26 & \textbf{0.94} \\ 
    2x256 & 1x128 & 513 & \textbf{1.92} & \textbf{0.2} & \textbf{2.62} & 0.9 \\ 
    2x128 & 1x64 & 142 & 1.28 & 0.31 & 1.58 & 0.68 \\ 
    2x64 & 1x32 & 42 & 1.34 & 0.34 & 1.18 & 0.43 \\ 
    \end{tabular}
    }
    \label{tab:architecture-comparison-librispeech}
    \vspace{-1.2em}
\end{table}

We are interested in the trade-off between compression speed, compression quality $q(\bm{p}, \mathcal{D}_f)$, and compression ratio $\frac{\left|\mathcal{D}_f\right|}{\left|\bm{p}\right|}$. We therefore evaluate SIREN and Siamese SIREN using small MLPs and over a small number of training iterations. We train each model for 2500 iterations, which results in a compression time of around 25 seconds per sample on one Titan RTX. We find that the first 2500 iterations have the biggest impact on reconstruction quality. Preliminary experiments conducted by training to 10k iterations had led to slightly better results, but with a clear trend of diminishing returns. Even though the underlying signal can be distinctly heard after a few hundred steps, it is challenging to remove the remaining background noise. Longer training times reduce the presence of the noise, but it is left clearly audible. Our proposed approach solves this by estimating and removing the noise.

To compress the network, we quantize the network weights with PTQ, which reduces the storage size by 4x. We also analyze how the performance scores react to drastic reductions in network size. We find that the reconstruction degrades heavily when the network does not have sufficient parameters to learn to fit the signal, as can be seen in \cref{tab:architecture-comparison-librispeech}.

We notice a significant gap in metric performance compared to subjective listening evaluations. This is a well-known problem in audio model evaluation \cite{7471749, kilgour2019frechet, Vinay2022EvaluatingGA}. We find that audio evaluation metrics tend to hold up better in speech signal analysis when compared to music signals.

\begin{figure}[t]
\includegraphics[width=\linewidth]{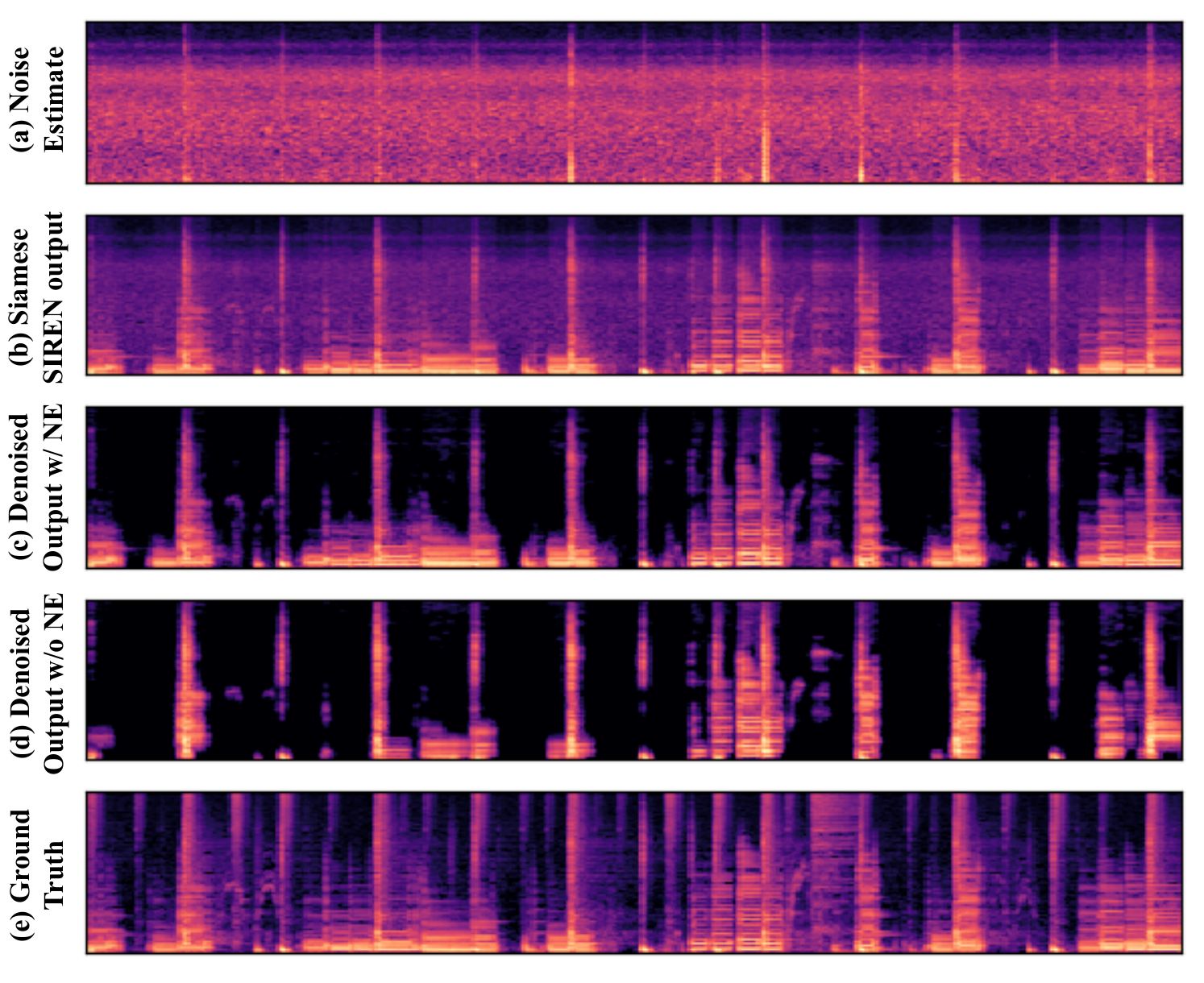}
\caption{Comparison of noise removal. We visualize noise estimate $\varepsilon_0$ (a) and signal $\hat{f}_0$ (b). We demonstrate denoising results with noise estimate (c) and without noise estimate (d), we observe better results when using a noise estimate.}
\label{fig:noise-reduce}
\end{figure}

To analyze the effect of estimating the noise distribution using our Siamese SIREN approach, we conduct an ablation study as shown in \cref{fig:noise-reduce}. We observe a more pronounced cut-off when no noise estimate is provided, especially for music signals. We also noticed that the largest discrepancy in signal fidelity comes from reconstruction -- the signal is subjectively only slightly more degraded after quantization.

\cref{tab:inr-comparison-librispeech} shows the results of our ablation study between the original SIREN and our Siamese SIREN. We observe the lowest score for the original SIREN, as the model cannot reconstruct the output after weight quantization. This can be seen in \cref{fig:training-comparison}.

We also measure the impact of increasing the proportion of shared layers (cf. \cref{tab:bonding-comparison-librispeech}). Unsurprisingly, we find that the parameter count has a large influence on reconstruction quality of the signal, indicating that there is a trade-off between reducing network size and maintaining reconstruction quality. Our experiments further show that keeping large parts of the network shared and only splitting the last layer into siamese heads achieves the best quality-size trade-off.

\begin{table}[t]
    \centering
    \caption{Evaluating the effect of layer sharing over random LibriSpeech samples.}
    \resizebox{\linewidth}{!}{%
    \begin{tabular}{ccccccc}
    \textbf{Shared} & \textbf{Siamese} & \textbf{\#Params ($10^3$)} & \textbf{ViSQOL}$\uparrow$ & \textbf{CDPAM}$\downarrow$ & \textbf{PESQ}$\uparrow$ & \textbf{STOI}$\uparrow$ \\
    \hline
    3x256 & 0     & 843 & 2.06          & 0.34         & 2.21          & \textbf{0.95} \\ 
    2x256 & 1x128 & 513 & \textbf{2.22} & \textbf{0.14} & \textbf{2.73} & 0.93 \\ 
    1x256 & 2x128  & 151 & 1.94          & 0.24         & 2.00          & 0.87 \\ 
    0 & 3x128   & 75  & 1.85          & 0.31         & 1.77          & 0.81 \\ 
    \end{tabular}
    }
    \label{tab:bonding-comparison-librispeech}
\end{table}

In summary, we present a first approach to audio compression using INRs. We introduce Siamese SIREN -- an extension to SIREN designed for audio compression and denoising tasks -- and find it to be a viable candidate for INR-driven compression of audio. We hope our work will help to facilitate future research on INRs for sound and speech.

\bibliography{example_paper}

\begin{thebibliography}{24}
\providecommand{\natexlab}[1]{#1}
\providecommand{\url}[1]{\texttt{#1}}
\expandafter\ifx\csname urlstyle\endcsname\relax
  \providecommand{\doi}[1]{doi: #1}\else
  \providecommand{\doi}{doi: \begingroup \urlstyle{rm}\Url}\fi

\bibitem[Benbarka et~al.(2021)Benbarka, H{\"o}fer, ul~Moqeet~Riaz, and
  Zell]{Benbarka2021SeeingIN}
Benbarka, N., H{\"o}fer, T., ul~Moqeet~Riaz, H., and Zell, A.
\newblock Seeing implicit neural representations as fourier series.
\newblock \emph{2022 IEEE/CVF Winter Conference on Applications of Computer
  Vision (WACV)}, pp.\  2283--2292, 2021.

\bibitem[Cartwright et~al.(2016)Cartwright, Pardo, Mysore, and
  Hoffman]{7471749}
Cartwright, M., Pardo, B., Mysore, G.~J., and Hoffman, M.
\newblock Fast and easy crowdsourced perceptual audio evaluation.
\newblock In \emph{2016 IEEE International Conference on Acoustics, Speech and
  Signal Processing (ICASSP)}, pp.\  619--623, 2016.
\newblock \doi{10.1109/ICASSP.2016.7471749}.

\bibitem[Chen et~al.(2021)Chen, Liu, and Wang]{chen2021learning}
Chen, Y., Liu, S., and Wang, X.
\newblock Learning continuous image representation with local implicit image
  function.
\newblock In \emph{Proceedings of the IEEE/CVF conference on computer vision
  and pattern recognition}, pp.\  8628--8638, 2021.

\bibitem[Chinen et~al.(2020)Chinen, Lim, Skoglund, Gureev, O'Gorman, and
  Hines]{chinen2020visqol}
Chinen, M., Lim, F. S.~C., Skoglund, J., Gureev, N., O'Gorman, F., and Hines,
  A.
\newblock Visqol v3: An open source production ready objective speech and audio
  metric, 2020.

\bibitem[Dupont et~al.(2021)Dupont, Goliński, Alizadeh, Teh, and
  Doucet]{dupont2021coin}
Dupont, E., Goliński, A., Alizadeh, M., Teh, Y.~W., and Doucet, A.
\newblock Coin: Compression with implicit neural representations, 2021.

\bibitem[Dupont et~al.(2022)Dupont, Loya, Alizadeh, Goli{\'n}ski, Teh, and
  Doucet]{dupont2022coin++}
Dupont, E., Loya, H., Alizadeh, M., Goli{\'n}ski, A., Teh, Y.~W., and Doucet,
  A.
\newblock Coin++: Data agnostic neural compression.
\newblock \emph{arXiv preprint arXiv:2201.12904}, 2022.

\bibitem[Kilgour et~al.(2019)Kilgour, Zuluaga, Roblek, and
  Sharifi]{kilgour2019frechet}
Kilgour, K., Zuluaga, M., Roblek, D., and Sharifi, M.
\newblock Fr\'echet audio distance: A metric for evaluating music enhancement
  algorithms, 2019.

\bibitem[Kingma \& Ba(2017)Kingma and Ba]{kingma2017adam}
Kingma, D.~P. and Ba, J.
\newblock Adam: A method for stochastic optimization, 2017.

\bibitem[Manocha et~al.(2021)Manocha, Jin, Zhang, and
  Finkelstein]{manocha2021cdpam}
Manocha, P., Jin, Z., Zhang, R., and Finkelstein, A.
\newblock Cdpam: Contrastive learning for perceptual audio similarity, 2021.

\bibitem[Mescheder et~al.(2019)Mescheder, Oechsle, Niemeyer, Nowozin, and
  Geiger]{mescheder2019occupancy}
Mescheder, L., Oechsle, M., Niemeyer, M., Nowozin, S., and Geiger, A.
\newblock Occupancy networks: Learning 3d reconstruction in function space.
\newblock In \emph{Proceedings of the IEEE/CVF conference on computer vision
  and pattern recognition}, pp.\  4460--4470, 2019.

\bibitem[Mildenhall et~al.(2021)Mildenhall, Srinivasan, Tancik, Barron,
  Ramamoorthi, and Ng]{mildenhall2021nerf}
Mildenhall, B., Srinivasan, P.~P., Tancik, M., Barron, J.~T., Ramamoorthi, R.,
  and Ng, R.
\newblock Nerf: Representing scenes as neural radiance fields for view
  synthesis.
\newblock \emph{Communications of the ACM}, 65\penalty0 (1):\penalty0 99--106,
  2021.

\bibitem[Panayotov et~al.(2015)Panayotov, Chen, Povey, and
  Khudanpur]{panayotov2015librispeech}
Panayotov, V., Chen, G., Povey, D., and Khudanpur, S.
\newblock Librispeech: an asr corpus based on public domain audio books.
\newblock In \emph{Acoustics, Speech and Signal Processing (ICASSP), 2015 IEEE
  International Conference on}, pp.\  5206--5210. IEEE, 2015.

\bibitem[Park et~al.(2019)Park, Florence, Straub, Newcombe, and
  Lovegrove]{Park2019DeepSDFLC}
Park, J.~J., Florence, P.~R., Straub, J., Newcombe, R.~A., and Lovegrove, S.
\newblock Deepsdf: Learning continuous signed distance functions for shape
  representation.
\newblock \emph{2019 IEEE/CVF Conference on Computer Vision and Pattern
  Recognition (CVPR)}, pp.\  165--174, 2019.

\bibitem[Rix et~al.(2001)Rix, Beerends, Hollier, and
  Hekstra]{Rix2001PerceptualEO}
Rix, A.~W., Beerends, J.~G., Hollier, M., and Hekstra, A.~P.
\newblock Perceptual evaluation of speech quality (pesq)-a new method for
  speech quality assessment of telephone networks and codecs.
\newblock \emph{2001 IEEE International Conference on Acoustics, Speech, and
  Signal Processing. Proceedings (Cat. No.01CH37221)}, 2:\penalty0 749--752
  vol.2, 2001.

\bibitem[Sainburg(2019)]{tim_sainburg_2019_3243139}
Sainburg, T.
\newblock timsainb/noisereduce: v1.0, June 2019.
\newblock URL \url{https://doi.org/10.5281/zenodo.3243139}.

\bibitem[Sitzmann et~al.(2020)Sitzmann, Martel, Bergman, Lindell, and
  Wetzstein]{sitzmann2020implicit}
Sitzmann, V., Martel, J., Bergman, A., Lindell, D., and Wetzstein, G.
\newblock Implicit neural representations with periodic activation functions.
\newblock \emph{Advances in Neural Information Processing Systems},
  33:\penalty0 7462--7473, 2020.

\bibitem[Strümpler et~al.(2022)Strümpler, Postels, Yang, van Gool, and
  Tombari]{strümpler2022implicit}
Strümpler, Y., Postels, J., Yang, R., van Gool, L., and Tombari, F.
\newblock Implicit neural representations for image compression, 2022.

\bibitem[Szatkowski et~al.(2022)Szatkowski, Piczak, Spurek, Tabor, and
  Trzciński]{szatkowski2022hypersound}
Szatkowski, F., Piczak, K.~J., Spurek, P., Tabor, J., and Trzciński, T.
\newblock Hypersound: Generating implicit neural representations of audio
  signals with hypernetworks, 2022.

\bibitem[Taal et~al.(2011)Taal, Hendriks, Heusdens, and Jensen]{Taal2011AnAF}
Taal, C.~H., Hendriks, R.~C., Heusdens, R., and Jensen, J.~R.
\newblock An algorithm for intelligibility prediction of time–frequency
  weighted noisy speech.
\newblock \emph{IEEE Transactions on Audio, Speech, and Language Processing},
  19:\penalty0 2125--2136, 2011.

\bibitem[Tancik et~al.(2020)Tancik, Srinivasan, Mildenhall, Fridovich-Keil,
  Raghavan, Singhal, Ramamoorthi, Barron, and Ng]{tancik2020fourier}
Tancik, M., Srinivasan, P.~P., Mildenhall, B., Fridovich-Keil, S., Raghavan,
  N., Singhal, U., Ramamoorthi, R., Barron, J.~T., and Ng, R.
\newblock Fourier features let networks learn high frequency functions in low
  dimensional domains, 2020.

\bibitem[Tzanetakis et~al.(2001)Tzanetakis, Essl, and
  Cook]{tzanetakis_essl_cook_2001}
Tzanetakis, G., Essl, G., and Cook, P.
\newblock Automatic musical genre classification of audio signals, 2001.
\newblock URL \url{http://ismir2001.ismir.net/pdf/tzanetakis.pdf}.

\bibitem[Vinay \& Lerch(2022)Vinay and Lerch]{Vinay2022EvaluatingGA}
Vinay, A. and Lerch, A.
\newblock Evaluating generative audio systems and their metrics.
\newblock In \emph{International Society for Music Information Retrieval
  Conference}, 2022.

\bibitem[Yu et~al.(2020)Yu, Ye, Tancik, and Kanazawa]{Yu2020pixelNeRFNR}
Yu, A., Ye, V., Tancik, M., and Kanazawa, A.
\newblock pixelnerf: Neural radiance fields from one or few images.
\newblock \emph{2021 IEEE/CVF Conference on Computer Vision and Pattern
  Recognition (CVPR)}, pp.\  4576--4585, 2020.

\bibitem[Zuiderveld et~al.(2021)Zuiderveld, Federici, and
  Bekkers]{zuiderveld2021lightweight}
Zuiderveld, J., Federici, M., and Bekkers, E.~J.
\newblock Towards lightweight controllable audio synthesis with conditional
  implicit neural representations, 2021.

\end{thebibliography}
\bibliographystyle{icml2023}

\newpage
\appendix
\onecolumn
\section{Noise Perception}
\label{appendix:noise-perception}
\begin{figure*}[b!]
\includegraphics[width=\textwidth]{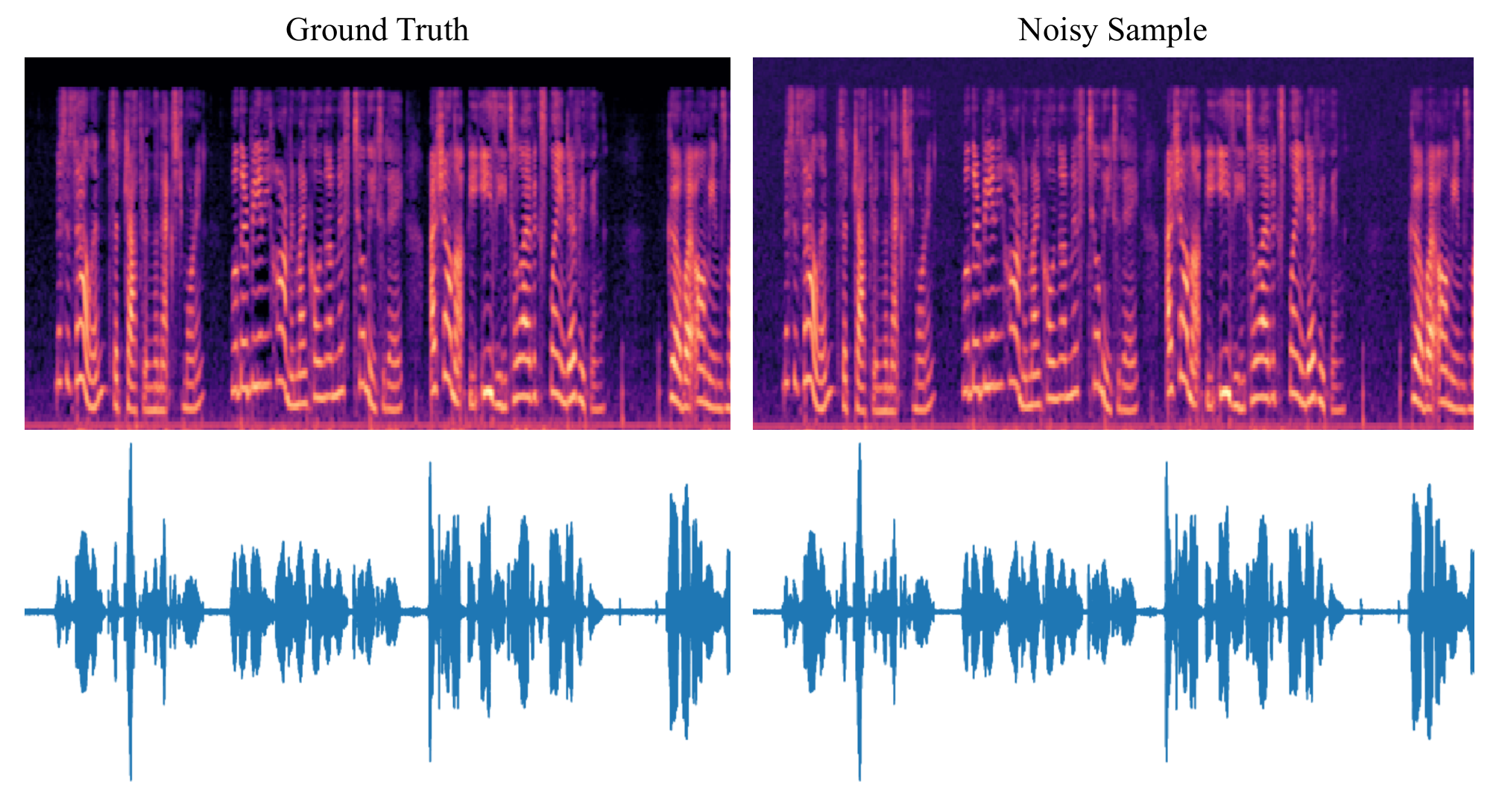}
\caption{Log-mel spectrogram of a LibriSpeech sample and the same sample with added $\varepsilon \sim \mathcal{N}(0,10^{-3}\mathbf I)$ noise. We can clearly see the difference in the spectrogram, but not in the waveform. This discrepancy is at the root of the challenge to remove noise. Even if the waveform is closely approximated, small errors are magnified and lead to distinctly audible noise.}
\label{fig:noise-human}
\end{figure*}

We demonstrate the effect of logarithmic hearing perception by adding noise $\varepsilon \sim \mathcal{N}(0,10^{-3}\mathbf I)$ to a LibriSpeech sample (cr. \cref{fig:noise-human}). We visualize this effect using log-mel spectrograms. The log-mel spectrogram is a perceptually-relevant amplitude and frequency representation of an audio sample. We observe a strong distinction between the ground truth log-mel spectrogram and the log-mel spectrogram of the noisy sample. However, this difference is not clearly noticeable when examining the amplitude waveform. Since we train SIREN and its extensions to learn on the amplitude waveform, it is unsurprising that there exists an inherent difficulty in removing noise.

\section{Model Training Parameters}
\label{appendix:training}
We train all models for 2500 iterations using Adam optimizer \cite{kingma2017adam} with the $\beta_1$ and $\beta_2$ parameters as proposed in the paper, learning rate of $1e^{-4}$, a weight decay of $1e^{-5}$ , and mean squared error as the loss function. We use frequency scaling with $\omega=100$, which we found to give better results compared to the widely used $\omega=30$. We use $L=16$ positional frequency encoding with scaling $\sigma=2$, resulting in a 33-dimensional input embedding which is passed into the MLP. Furthermore, we normalize time inputs into the range of $\mathcal{T} = \left[-1,1\right]$. In our early experiments we tested other loss functions, scaling and positional frequencies, optimizers, learning rate and weight decay values, and learning rate schedules, but found them to have little effect on reconstruction quality. Furthermore, the \textit{noise reduce} algorithm we used allows to only remove a percentage of detected noise. We did not use this feature, as we found keeping partial noise did not increase reconstruction quality.

\newpage
\section{SIREN and Stationary Background Noise}
\label{appendix:siren-audio}
\begin{figure*}[b!]
\centering
\includegraphics[scale=0.73]{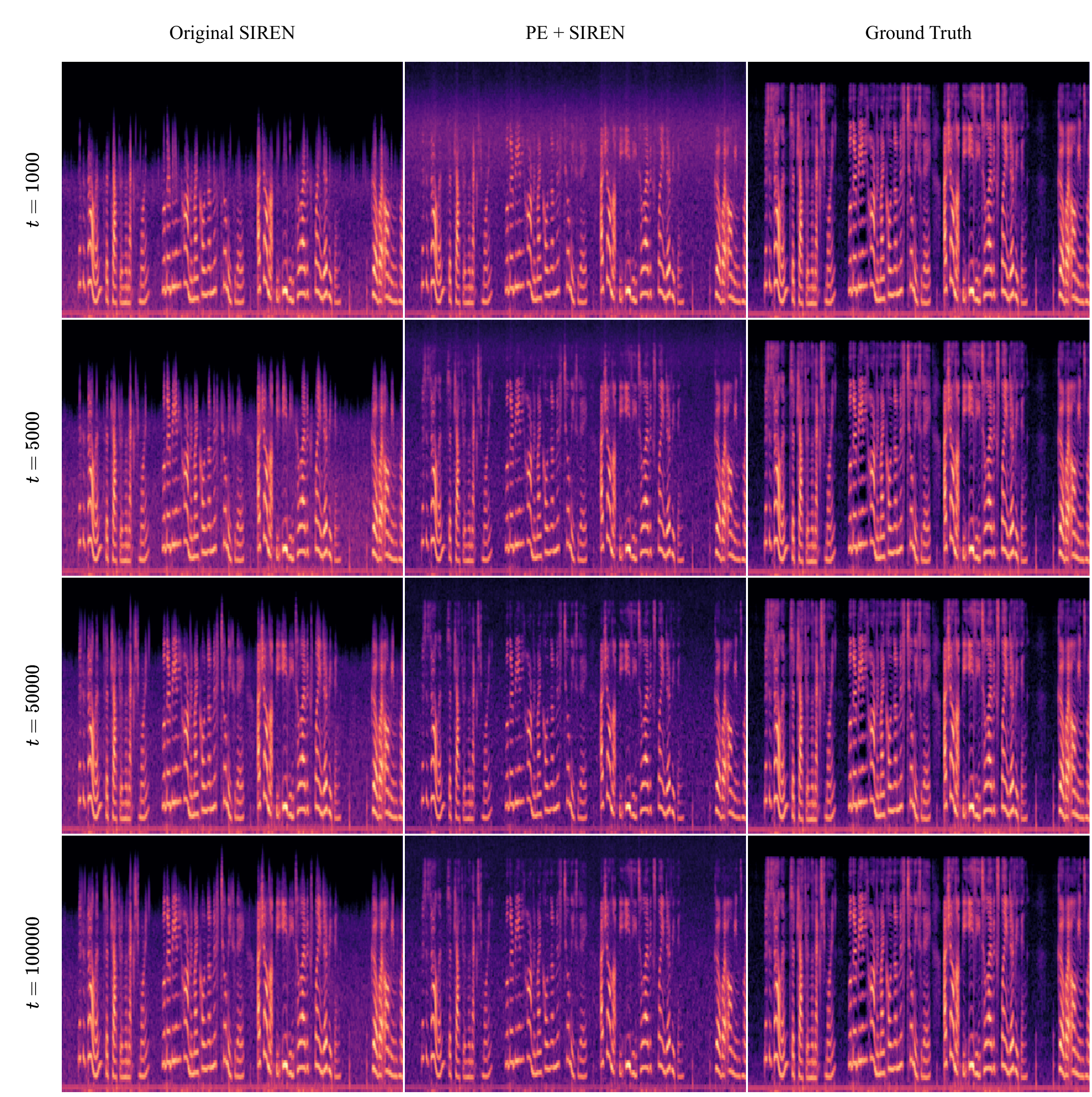}
\caption{Log-mel spectrogram of unquantized original SIREN and unquantized SIREN with positional encoding over 100k iterations. We observe significant background noise in both reconstructions.}
\label{fig:unquantized-siren}
\end{figure*}

We demonstrate the challenge of removing noise from an audio reconstruction. We train the original SIREN setup for audio, as described by \cite{sitzmann2020implicit}, and SIREN with positional encoding, over 100k iterations on a LibriSpeech sample (cf. \cref{fig:unquantized-siren}). For this experiment we do not quantize model weights. Comparing the spectrograms, we notice that without positional encoding the original SIREN struggles to reproduce high-frequency bands while simultaneously containing substantial stationary background noise in the mid-frequency and low-frequency bands. SIREN with positional encoding is capable of learning high-frequency content, however, it also produces substantial noise in these high-frequency bands. Furthermore, we observe only minimal improvement when training for more than 5000 iterations. 

\end{document}